\documentclass[sigconf,review=false]{acmart} 

\copyrightyear{2025} 
\acmYear{2025} 
\setcopyright{acmlicensed}\acmConference[WWW '25]{Proceedings of the ACM Web Conference 2025}{April 28-May 2, 2025}{Sydney, NSW, Australia}
\acmBooktitle{Proceedings of the ACM Web Conference 2025 (WWW '25), April 28-May 2, 2025, Sydney, NSW, Australia}
\acmDOI{10.1145/3696410.3714600}
\acmISBN{979-8-4007-1274-6/25/04}

\usepackage{enumitem}
\usepackage{booktabs}  
\usepackage{multirow}  
\usepackage{graphicx}  
\usepackage{amsmath}  
\usepackage{mathrsfs}  
\usepackage{algorithmic}
\usepackage{algorithm}
\usepackage{array}
\begin{document}

\title{ESANS: Effective and Semantic-Aware Negative Sampling for Large-Scale Retrieval Systems}

\author{Haibo Xing}
\orcid{0009-0006-5786-7627}
\affiliation{%
  \institution{Alibaba International Digital Commerce Group}
  \city{Hangzhou}
  \state{Zhejiang}
  \country{China}
}

\author{Kanefumi Matsuyama}
\orcid{0009-0002-1365-5375}
\affiliation{%
  \institution{Alibaba International Digital Commerce Group}
  \city{Hangzhou}
  \state{Zhejiang}
  \country{China}
}

\author{Hao Deng}
\orcid{0009-0002-6335-7405}
\affiliation{%
  \institution{Alibaba International Digital Commerce Group}
  \city{Beijing}
  \state{Beijing}
  \country{China}
}

\author{Jinxin Hu}
\authornote{Corresponding Author}
\orcid{0000-0002-7252-5207}
\affiliation{%
  \institution{Alibaba International Digital Commerce Group}
  \city{Beijing}
  \state{Beijing}
  \country{China}
}

\author{Yu Zhang}
\orcid{0000-0002-6057-7886}
\affiliation{%
  \institution{Alibaba International Digital Commerce Group}
  \city{Beijing}
  \state{Beijing}
  \country{China}
}

\author{Xiaoyi Zeng}
\orcid{0000-0002-3742-4910}
\affiliation{%
  \institution{Alibaba International Digital Commerce Group}
  \city{Hangzhou}
  \state{Zhejiang}
  \country{China}
}

\renewcommand{\shortauthors}{Haibo Xing et al.}

\begin{abstract}
  Industrial recommendation systems typically involve a two-stage process: retrieval and ranking, which aims to match users with millions of items. In the retrieval stage, classic embedding-based retrieval (EBR) methods depend on effective negative sampling techniques to enhance both performance and efficiency. However, existing techniques often suffer from false negatives, high cost for ensuring sampling quality and semantic information deficiency. To address these limitations, we propose Effective and Semantic-Aware Negative Sampling (ESANS), which integrates two key components: Effective Dense Interpolation Strategy (EDIS) and Multimodal Semantic-Aware Clustering (MSAC). EDIS generates virtual samples within the low-dimensional embedding space to improve the diversity and density of the sampling distribution while minimizing computational costs. MSAC refines the negative sampling distribution by hierarchically clustering item representations based on multimodal information (visual, textual, behavioral), ensuring semantic consistency and reducing false negatives. Extensive offline and online experiments demonstrate the superior efficiency and performance of ESANS.
\end{abstract}

\ccsdesc[500]{Information systems~Retrieval models and ranking}

\keywords{Recommendation systems, Embedding-based retrieval, Negative sampling}



\maketitle

\section{Introduction}
\label{sec:intro}
Recommendation systems have been widely adopted across diverse domains, including online e-commerce, advertising, short video platforms and delivery services\ \citep{fan2021attacking, fan2020graph, zhao2021autoloss}, owing to their effectiveness in mitigating information overload by providing tailored recommendations from large-scale item collections\ \citep{dscf, fan2022comprehensive}. Industrial recommendation systems typically involve two stages: retrieval and ranking. The retrieval stage is responsible for retrieving thousands of candidate items, whereas the ranking stage predicts the likelihood of user interaction with these candidates. Considering that retrieval tasks can be formulated as identifying the nearest neighbors in a vector space, substantial research has been devoted to developing high-quality representations for both users and items. Collaborative Filtering (CF) methods\ \citep{sarwar2001item, chen2012svdfeature, hu2008collaborative, rendle2010factorization} address this issue by encoding user preference and item representation into low-dimensional embedding space, based on historical interacted information. With the rapid development of deep learning, neural networks have been widely adopted in personalized recommendation systems\ \citep{chen2017attentive, ge2020graph, wei2019mmgcn}. Recently, Embedding-Based Retrieval (EBR) methods\ \citep{covington2016deep, cen2020controllable, li2019multi} have demonstrated significantly better performance compared to traditional CF methods, establishing themselves as the dominant approach in recommendation systems. EBR methods encode user and item information into separate embeddings using parallel neural networks, and these embeddings are trained through the strategy of contrastive learning\ \citep{fan2019deep, rendle2012bpr, mikolov2013distributed}. 

\begin{figure}
  \includegraphics[width=0.48\textwidth]{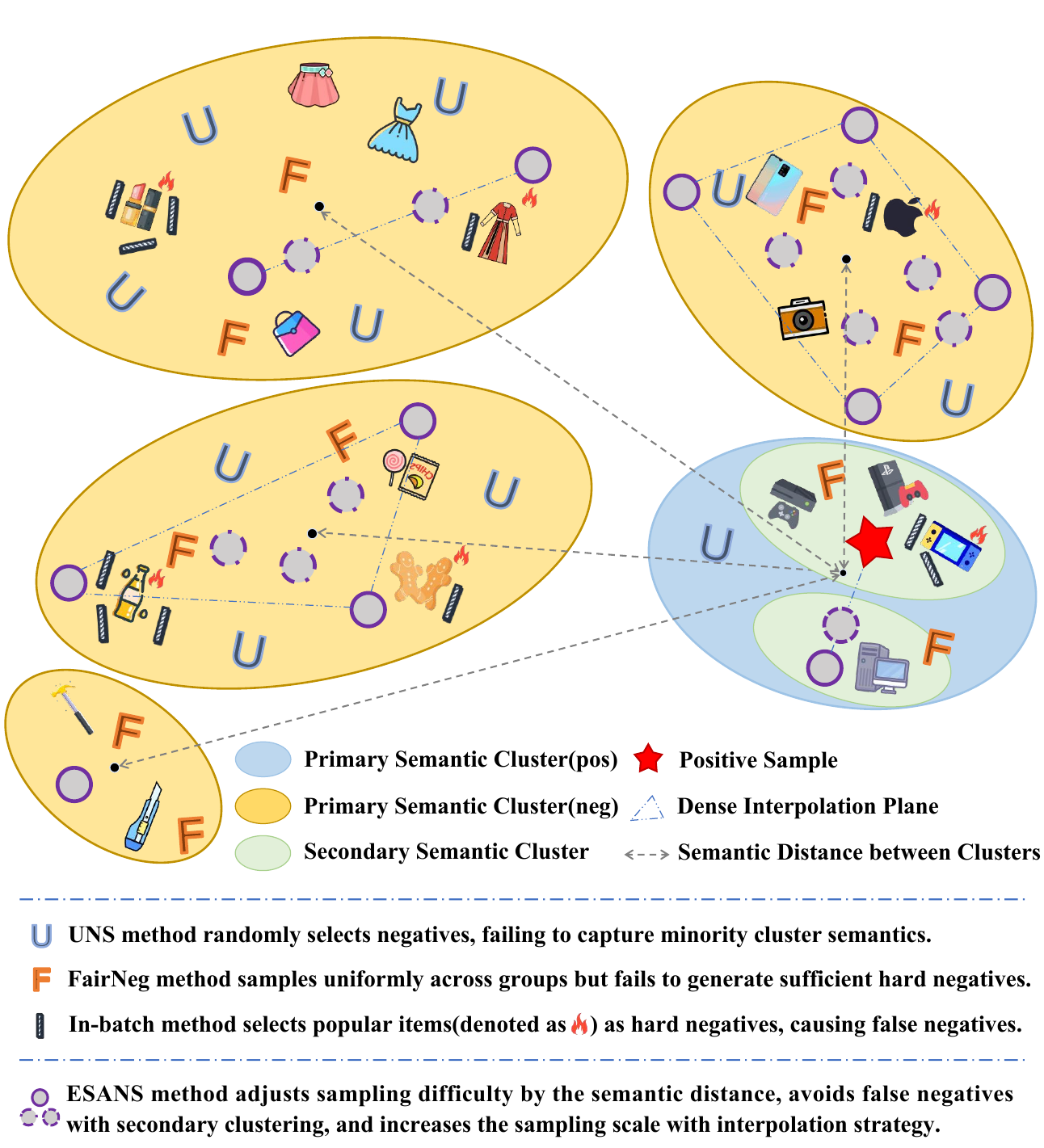}
  \caption{Visual diagram of our ESANS compared with other methods. Each method has sampled ten negatives equally.}
  \Description{xx.}
  \label{fig:intro}
\vspace{-3pt}
\end{figure}

EBR methods rely heavily on the contrast between positive and negative samples to produce distinguishable representations. The careful selection of negatives is crucial to enhancing the model’s ability to differentiate between relevant and irrelevant items, significantly impacting overall retrieval performance. The classic Uniform Negative Sampling (UNS) method\ \citep{rendle2012bpr, he2017neural} randomly selects negatives from the item candidate set, providing efficiency but yielding \textbf{low-quality samples}. Following this, additive margin\ \citep{wang2018additive} and temperature coefficient\ \citep{wang2021understanding, lou2022re} adjust the contrastive loss function to mine high-quality negatives from naive negatives sampled by UNS. FairNeg\ \citep{chen2023fairly} reweights negatives in accordence with item group fairness to provide high-quality samples. Adap-$\tau$\ \citep{chen2023adap} dynamically adjusts the temperature coefficient for reweighting uniform negatives in accordence with their relevance to user interests. However, these methods fail to introduce more challenging negatives and further expand the scale of sampling which limit the performance of EBR methods. To address this issue, In-batch sampling\ \citep{chen2017sampling} introduces relatively harder negatives by the in-batch sharing strategy. Airbnb\ \citep{grbovic2018real} heuristically introduces orders from the same city as harder negatives. MixGCF\ \citep{huang2021mixgcf} employs a hop-mixing interpolation technique in Graph Neural Networks (GNNs) to generate virtual hard negatives. However, these methods fail to effectively adjust the difficulty of negatives and distinguish users' potential interests from hard negatives, which may exacerbate the issue of \textbf{false negatives} (i.e. items relevant to users’ potential interests but incorrectly regarded as negatives). Moreover, existing methods require substantial computational resources to further \textbf{improve the sampling quality} (i.e. sufficient hard negatives)\ \citep{chen2020simple}. From the contrastive learning perspective, these methods are unable to regulate sampling strategies based on \textbf{semantic information} in the real world, rendering the sampling process a black box. To address these issues, we redesign the sampling space from a multimodal perspective and propose a controllable negative sampling algorithm based on well-defined semantic distance.

Specifically, Inspired by recent works in multi-modal learning\ \citep{radford2021learning,multimodalreview} and vector quantization techniques\ \citep{van2017neural}, we propose the \textbf{E}ffective and  \textbf{S}emantical-\textbf{A}ware \textbf{N}egative \textbf{S}ampling (ESANS) to address these challenges in the sampling process. Our method consists of two main components: the first part is Effective Dense Interpolation Strategy (EDIS), and the second part is Multimodal Semantic-Aware Clustering (MSAC). EDIS is devised to generate a sufficient number of virtual samples within the low-dimensional embedding space. More specifically, generating virtual samples among existing negatives creates a more uniform, dense, and diverse sampling distribution. Virtual samples positioned between the positive sample and surrounding negative samples contribute to gradually enhance the discriminatory ability of the neural network. By adjusting the interpolation parameters and strategies, we can control the difficulty of generated negatives and generate sufficient hard negatives. Meanwhile, in contrast to memory banks\ \citep{he2020momentum}, interpolation within the low-dimensional embedding space leads to minimal computational cost and eliminates the need for extra memory storage. 

Nevertheless, EDIS strongly relies on the judicious selection of negative sample anchors. In practice, virtual negative samples generated via interpolation may lack clear semantic information, occasionally producing meaningless samples. For example, interpolating between "iPhone" and "Cola" produces meaningless results, potentially introducing noise. Moreover, interpolating among randomly sampled negative anchors may introduce false negatives, further complicating the training process. 

To address these deficiencies, we propose the MSAC method to optimize the sampling space by integrating the real-world semantic information. Firstly, we propose a multimodal-aligned technique to fuse multi-perspective item information from visual, textual and behavioral perspectives. Subsequently, a two-level vector quantized clustering approach is employed to assign semantic representations into multiple secondary clusters. Consequently, we can mitigate the issue of false negatives by selecting hard negatives from the same primary cluster as the positive sample, while ensuring they belong to a different secondary cluster. Additionally, we dynamically calibrate the sampling probabilities for each negative cluster to control the difficulty of negatives and refine the sampling quality. It is worth noting that this calibration is precisely guided by the semantic distance between the cluster centers of positives and negatives. This allows us to adjust the difficulty of the sampling process by increasing the sampling probabilities of clusters that are semantically similar to the positive cluster. Once the MSAC is introduced, EDIS based on semantics can be performed within the well-established semantic clusters. More specifically, we can ensure that the interpolated outcomes remain confined within the convex hull of that cluster. This intrinsic constraint preserves a measurable degree of semantic consistency and \textbf{"real-world applicability"} in the interpolated samples. Furthermore, interpolation between positives and hard negatives is also employed to generate additional high-quality hard negatives. Figure \ref{fig:intro} shows the comparison between our ESANS and other methods. Our contributions can be summarized as follows:

\begin{itemize}[noitemsep, topsep=0pt, leftmargin=*]
\item We propose a novel and effective sampling approach called ESANS, which provides explicit semantics guidance for interpolation negative sampling. Moreover, ESANS effectively enhances the diversity and richness of negative samples and allows for controllable negative sample difficulty, thereby boosting performance.
\item We propose a general multimodal-aligned clustering approach that captures the multi-perspective similarities among candidate items on e-commerce platforms, thereby enabling a more refined semantic description in the interpolation space and eliminating false negative instances in the hard negative sampling process.
\item We provide both extensive offline and online experiments to demonstrate the effectiveness and the efficiency of ESANS.
\end{itemize}

\section{Related Work}
\label{sec:related}
This section presents a brief review of the relevant literature, specifically addressing techniques for negatives re-weighting, heuristic negative sampling, and model-based negative sampling.

\noindent \textbf{Negatives Re-weighting.}
UNS\ \citep{rendle2012bpr, he2017neural} represents the foundational negative sampling method, where negative samples are uniformly drawn from the entire dataset. The simplicity of UNS’s algorithmic design provides substantial efficiency gains. Nevertheless, it exhibits notable deficiencies in the quality of negative samples. UMA2\ \citep{lou2022re} computes the sampling probabilities of random negative samples according to the current model and subsequently employs the Inverse Probability Weighting (IPW) technique to assign loss weights to these negative samples. The method proposed by\ \citep{rendle2014improving} implements position-weighted approach for negative samples, where the weight is determined by the sample's ranking position. These approaches mine high-quality negatives from naive negatives sampled by UNS, which, however, fails to introduce more challenging negatives.

\noindent \textbf{Heuristic Negative Sampling.}
Heuristic negative sampling algorithms primarily define the sampling distribution by predefined heuristic rules. Popularity-biased Negative Sampling (PNS) \citep{chen2017sampling} utilizes item popularity as the sampling probability. Airbnb \citep{grbovic2018real} applies personalized negative sampling within the same city, assuming bookings in the same location exhibit similar patterns. While this approach enhances the sampling process, it solely focuses on similarity-based sampling, neglecting sampling bias. CBNS \citep{wang2021cross} employs in-btch negative sampling and expands the negative sample set by incorporating previously trained items. The method\ \citep{yi2019sampling} incorporates estimated item frequency into the batch softmax cross-entropy loss to reduce sampling bias within the batch. MNS \citep{yang2020mixed} integrates UNS with in-batch negative sampling, adopting a hybrid strategy. While these methods enhance sampling quality, they introduce popularity bias, aggravating the Sample Selection Bias (SSB) issue. In contrast, our method enhances sampling quality via a multimodal-aligned clustering algorithm and dense interpolation negative sampling, while effectively mitigating sampling bias.

\noindent \textbf{Model-based Negative Sampling.}
Model-based negative sampling algorithms are highly effective at selecting high-quality negative samples. Model-based scoring methods are demonstrated by Dynamically Negative Sampling (DNS) \citep{zhang2013optimizing} and ESAM \citep{chen2020esam}, where the current model scores samples and selects the highest-scoring ones as negative samples. Adversarial learning methods also contribute to sampling improvements. 
MixGCF \citep{huang2021mixgcf} employs a hop-mixing technique to synthesize hard negative samples by leveraging the user-item graph structure and the aggregation mechanism of Graph Neural Networks (GNNs). IRGAN \citep{wang2017irgan} utilizes two recommendation models, a discriminator and a generator, trained adversarially. AdvIR \citep{park2019adversarial} and RNS \citep{ding2019reinforced} further optimize IRGAN's structure, improving both efficiency and performance. The Adap-$\tau$\ \citep{chen2023adap} adaptively adjusts the temperature coefficient of the loss function by calculating the loss for each user and the corresponding random negative samples. This method leverages personalized user preferences to effectively identify hard negative samples. FairNeg \citep{chen2023fairly} enhances the sampling distribution by fairly sampling from groups and then reweighting based on their relevance to the user. Our method precisely controls the difficulty of negatives, improving sampling quality and eliminating false negatives without increasing the complexity of the retrieval model.

\begin{figure*}[htbp]
  \includegraphics[width=\textwidth]{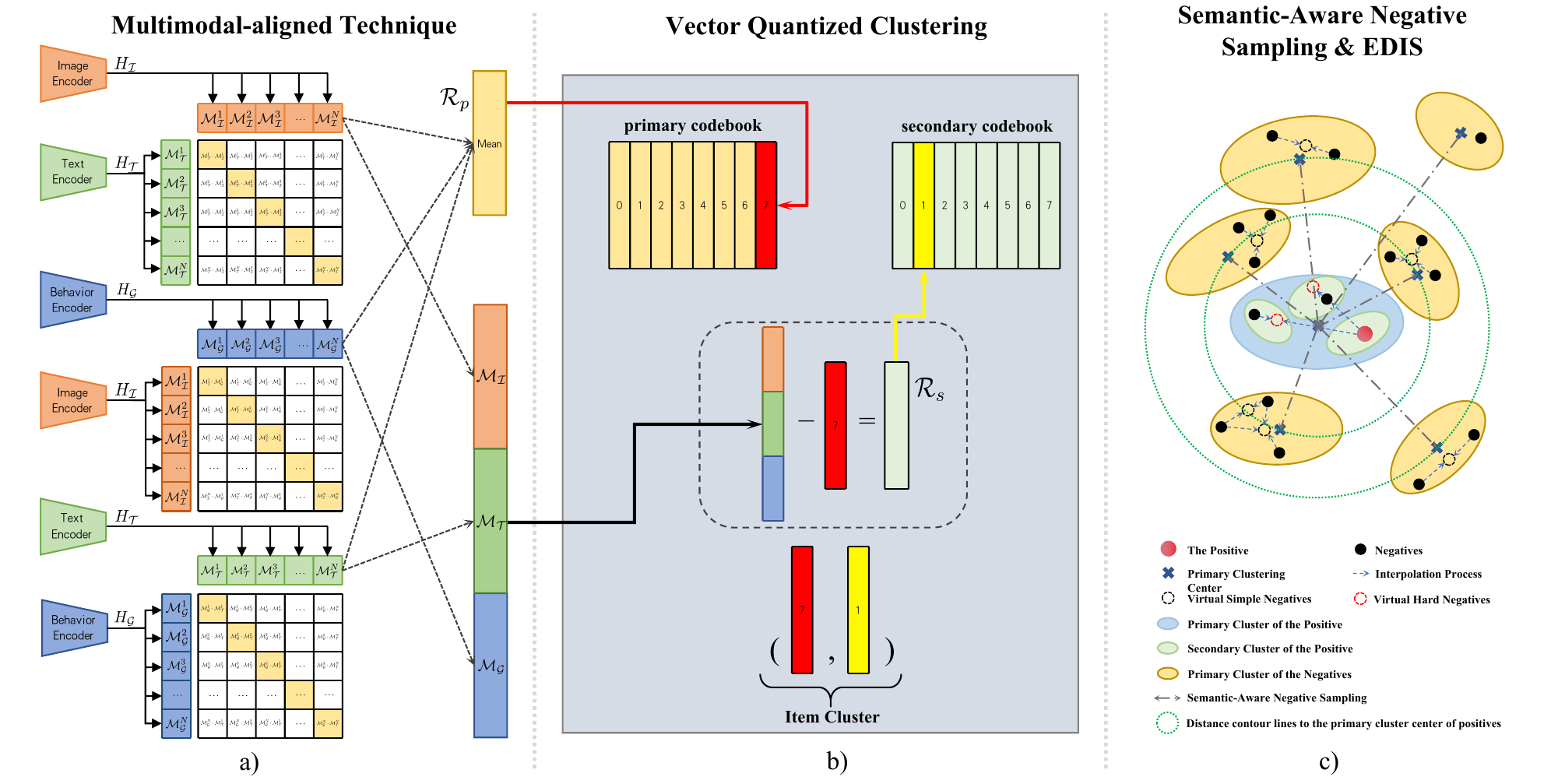}
  \caption{Our proposed ESANS framework. a) Multimodal-aligned Technique. b) Vector Quantized Clustering with Cascaded Codebooks. c) Semantic-Aware Negative Sampling \& Effective Dense Interpolation Strategy (EDIS).}
  \Description{xx.}
  \label{fig:frameworkb}
\end{figure*}

\section{Methodology}
\label{sec:method}
In this section, we formulate the problem and describe our proposed framework specifically, as well as introducing the detailed process of our negative sampling method.

\subsection{Problem Formulation}
The primary objective of the retrieval stage in industrial recommendation systems is to efficiently retrieve a potentially relevant subset of items from a large item pool \( \mathcal{I}\) for each user \( u \in \mathcal{U}\). In pursuit of this objective, each instance can be represented by a tuple \( (\mathcal {B}_u, \mathcal{P}_u,\mathcal{I}_i) \) where \(\mathcal {B}_u\) denotes the sequence of user historical behaviors, \( \mathcal{P}_u\) denotes the basic profile of user \(u\), \(\mathcal{I}_i\) denotes the information of target item such as item id and category id. In the classical two-tower architecture\ \citep{wang2018additive} of the EBR models, users and items are separated into two individual encoders to reduce online computational complexity. We can define the user encoder as \(f_{user}\) and the item encoder as \(g_{item}\), so we have:
\begin{equation}
\begin{split}
\mathbf{u}_u&={f_{user}(\mathcal{B}_u, \mathcal{P}_u) },\\
\mathbf{v}_i&={g_{item}(\mathcal{I}_i) },
\end{split}
\label{eq:uve}
\end{equation}
where \(\mathbf{u}_u \in \mathbb{R}^{d_k \times 1}\) is the output vector of the user encoder called user embedding, and \(\mathbf{v}_i \in \mathbb{R}^{d_k \times 1}\) is the output vector of the item encoder called item embedding. \(K\) denotes the dimension of output embeddings. Finally, the relevance of a user-item pair can be estimated by a scoring function:
\begin{equation}
s(\mathbf{u}, \mathbf{v}) = \mathbf{u}^\top \mathbf{v}.
\label{eq:score_function}
\end{equation}
\subsection{Overall Framework}
\label{sec:framework}
As previously discussed, existing methods fail to balance sampling quality, bias, and efficiency simultaneously. To address these limitations, we designed ESANS, as illustrated in Figure \ref{fig:frameworkb}. ESANS consists of two main components:

\begin{itemize}[noitemsep, topsep=0pt, leftmargin=*]
\item \textbf{Multimodal Semantic-Aware Clustering (MSAC)}, which performs hierarchical clustering based on visual, textual, and behavi-or based representations to optimize the sampling process by integrating semantic information. Our proposed method addresses the limitations of unclear anchor semantics, improves sampling quality, and reduces the risk of introducing false negatives.
\item \textbf{Effective Dense Interpolation Strategy (EDIS)}, which employs linear interpolation among existing samples within the same semantic cluster to make sure the semantic consistency. Our proposed method works with minimal computational cost, enhances the diversity and richness of negative samples, and facilitates the controllable difficulty of hard negative samples.
\end{itemize}

\subsection{Multimodal Semantic-Aware Clustering}
\label{sec:mc}
Most existing negative sampling methods ignore semantic correlations among samples. Against this deficiency, our MSAC is proposed to capture the multi-perspective similarities among items and incorporate explicit semantics into the negative sampling process.

\subsubsection{Multimodal-aligned Technique}

When users browse items on the e-commerce platform, they primarily perceive items through three views: visual images, descriptive text, and collaborative filtering recommendations. To generate a comprehensive description of items, it is necessary to consider these views concurrently. The visual representations \(\mathcal{R_I}\) and textual representations \(\mathcal{R_T}\) can be pretrained by individual specific encoders\ \citep{huang2020pixel,li2019visualbert}. The behavior-based representations \(\mathcal{R_G}\) can be pre-trained using graph representation learned based on a substantial number of user behaviors. \textbf{It is worth noting that modal embeddings can be generated in advance using existing pre-trained models and then kept frozen during the multimodal alignment process}.

Given a mini-batch of $N$ items, we design multimodal-aligned linear transformations for each view.
\begin{equation}
\begin{split}
\mathcal{M_I} &= H_\mathcal{I}(\mathcal{R_I}) \in \mathbb{R}^{N\times d_m}, \\
\mathcal{M_T}&=H_\mathcal{T}(\mathcal{R_T}) \in \mathbb{R}^{N\times d_m},\\
\mathcal{M_G}&=H_\mathcal{G}(\mathcal{R_G}) \in \mathbb{R}^{N\times d_m},
\end{split}
\label{eq:m}
\end{equation}
where \(H_*\) denotes the linear transformation of each view, \(\mathcal{ M_*} \) denotes the output embedding of each view, \(d_m\) denotes the output dimension of each view.

Inspired by the Contrastive Language-Image Pre-training (CLIP) \citep{radford2021learning}, We propose a multimodal alignment method to fuse item representations from three perspectives. Given a dataset of \(\mathcal{M_*}\) that consists of a collection of output embeddings \(\{\mathcal{M}^i_\mathcal{I}, \mathcal{M}^i_\mathcal{T}, \mathcal{M}^i_\mathcal{G}\}_{i=1}^N\), we contrast congruent and incongruent pairs across any two modalities. For instance, we sample from the joint distribution of image-text modals \( \mathbf{x}_{\mathcal{I} -\mathcal{ T}} \backsim \mathbf{P}(\mathcal{ M_I}, \mathcal{ M_T}) \) or \(\mathbf{x}_{\mathcal{ I} -\mathcal{ T}}=\{\mathcal{M}^i_\mathcal{I}, \mathcal{M}^i_\mathcal{T}\}\), which we call positive samples. We sample from the product of marginals, \(\mathbf{y}_{\mathcal{ I}- \mathcal{ T}} \backsim \mathbf{P}(\mathcal{ M_I})\mathbf{P}(\mathcal{ M_T})\) or \(\mathbf{y}_{\mathcal{ I} -\mathcal{ T}}=\{\mathcal{M}^i_\mathcal{I}, \mathcal{M}^j_\mathcal{T}\}\), which we call negative samples. Multimodal-aligned encoders are optimized to correctly select a single positive sample \(\mathbf{x}_{\mathcal{ I}- \mathcal{ T}}\) out of the set \(\mathcal{ S}=\{\mathbf{x}_{\mathcal{ I} -\mathcal{ T}},\mathbf{y}_{\mathcal{ I} -\mathcal{ T}}^1,...,\mathbf{y}_{\mathcal{ I} -\mathcal{ T}}^{N-1}\}\) which contains $N-1$ negative samples:
\begin{equation}
    \begin{split}
        \mathcal{ L}_{align}^{\mathcal{I}-\mathcal{T}}&=-\mathop{\mathbb{E}}\limits_{\mathcal{S}}[ \log \frac {h(\mathbf{x}_{\mathcal{ I}- \mathcal{ T}})} {h(\mathbf{x}_{\mathcal{ I}- \mathcal{ T}})+\sum_{i=1}^{N-1}{h(\mathbf{y}_{\mathcal{ I}- \mathcal{ T}}^i)}}], \\
        \mathcal{ L}_{align}^{\mathcal{I}-\mathcal{G}}&=-\mathop{\mathbb{E}}\limits_{\mathcal{S}}[ \log \frac {h(\mathbf{x}_{\mathcal{ I}- \mathcal{ G}})} {h(\mathbf{x}_{\mathcal{ I}- \mathcal{ G}})+\sum_{i=1}^{N-1}{h(\mathbf{y}_{\mathcal{ I}- \mathcal{ G}}^i)}}], \\
        \mathcal{ L}_{align}^{\mathcal{G}-\mathcal{T}}&=-\mathop{\mathbb{E}}\limits_{\mathcal{S}}[ \log \frac {h(\mathbf{x}_{\mathcal{ G}- \mathcal{ T}})} {h(\mathbf{x}_{\mathcal{ G}- \mathcal{ T}})+\sum_{i=1}^{N-1}{h(\mathbf{y}_{\mathcal{ G}- \mathcal{ T}}^i)}}], 
    \end{split}
    \label{eq:xx}
\end{equation}
where \(h (\cdot)\) is the cosine similarity operation after exponentiation, \(\mathcal{ L}_{align}^{\mathcal{I}-\mathcal{T}}\) is the alignment loss between visual and textual modals, \(\mathcal{ L}_{align}^{\mathcal{I}-\mathcal{G}}\) is the alignment loss between visual and behavior-based modals, \(\mathcal{ L}_{align}^{\mathcal{G}-\mathcal{T}}\) is the alignment loss between behavior-based and textual modals. 

\subsubsection{Vector Quantized Clustering with Cascaded Codebooks}

While aligning \(\mathcal {M_I}, \mathcal {M_T}, \mathcal {M_G}\) into the same embedding space, we simultaneously quantize these representations into several clusters with cascaded codebooks, as illustrated in Figure \ref{fig:frameworkb}. Specifically, the primary codebook is designed to effectively differentiate coarse-level item representations, while the secondary codebook enhances this distinction by refining the differentiation of fine-grained item representations, especially when significant disparities persist among aligned representations across partial modalities.

The \textit{\textbf{primary codebook}}  \(C_p=\{z_p^k\}_{k=1}^{K_p}\) consists of \({K_p}\) codewords\ \citep{lee2022autoregressive} and the dimension of each codeword is \(d_m\). The clustering stage is conducted by calculating the mean of the aligned embeddings:
\begin{equation}
\mathcal {R}_p^i=\frac {1} {3}(\mathcal{M}^i_\mathcal{I}+\mathcal{M}^i_\mathcal{T}+\mathcal{M}^i_\mathcal{G}).
\label{eq:mean}
\end{equation}

Subsequently \(\mathcal{ R}_p = \{\mathcal{R}_p^i\}_{i=1}^N\) is quantized by assigning it to the nearest codeword within the primary codebook. We denote that the nearest codeword to \(\mathcal{ R}_p^i\) is \(C_p^i=\arg \min_k \|\mathcal{ R}_p^i-z_p^k \|\). 

In the \textit{\textbf{secondary codebook}}, we compute the residual between \(\{\mathcal {M_I},\mathcal{ M_T},\mathcal{ M_G}\}\) and the primary corresponding codeword \(z_p^{C_p^i}\). These residuals are concatenated to a vector \(\mathcal{ R}_s^i\), which is used to describe the modal-specific information between different items.
\begin{equation}
\mathcal{ R}_s^i=[\mathcal{M}^i_\mathcal{I}-z_p^{C_p^i};\mathcal{M}^i_\mathcal{T}-z_p^{C_p^i};\mathcal{M}^i_\mathcal{G}-z_p^{C_p^i}].
\label{eq:residual}
\vspace{-10pt}
\end{equation}
\begin{figure}
  \includegraphics[width=0.45\textwidth]{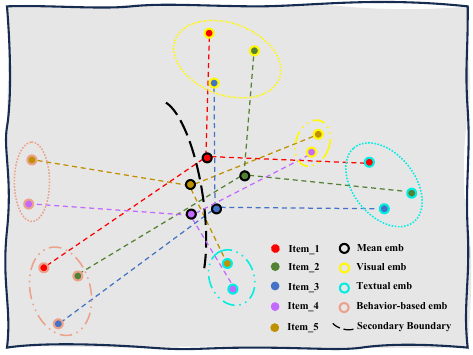}
  \caption{The visualization of items in the representation space during secondary clustering. Although item 1-3 and item 4-5 have similar mean embeddings, but in each view thier embeddings differ significantly, resulting in their assignment to different secondary clusters. By leveraging three modalities, clustering accuracy is significantly enhanced.}
  \Description{xx.}
  \label{fig:secondary}
  \vspace{-15pt}
\end{figure}

The advantages of using information from three modalities for secondary clustering are illustrated in Figure \ref{fig:secondary}. Similar to the primary clustering, we select the codeword closest to \(\mathcal{ R}_s= \{\mathcal{R}_s^i\}_{i=1}^N\) from another codebook \(C_s=\{z_s^k\}_{k=1}^{K_s}\), where \({K_s}\) denotes the number of codewords in the secondary codebook. The nearest secondary codeword to \(\mathcal{ R}_s^i\) is recorded as \(C_s^i=\arg \min_k \|\mathcal{ R}_s^i-z_s^k \|\).

Once we have all cluster indice for an item, the clustering loss can be defined as:
\begin{equation}
\mathcal{L}_{\text{SQ}} = \sum_{i=1}^N \|\mathcal{ R}_p^i-z_p^{C_p^i} \|^2+ \sum_{i=1}^N \|\mathcal{ R}_s^i-z_s^{C_s^i} \|^2.
\label{eq:codebook}
\end{equation}

We properly initialize \(C_p\) and \(C_s\) with the $k-means$ algorithm to avoid the codebook collapse. Finally, the loss function for multimodal-aligned clustering is given by Equation \ref{eq:total}:
\begin{equation}
\mathcal{ L}=\beta_1\mathcal{ L}_{align}^{\mathcal{I}-\mathcal{T}}+\beta_2\mathcal{ L}_{align}^{\mathcal{I}-\mathcal{G}}+\beta_3\mathcal{ L}_{align}^{\mathcal{G}-\mathcal{T}}+\mathcal{L}_{\text{SQ}}. 
\label{eq:total}
\end{equation}

\subsubsection{Semantic-Aware Negative Sampling}

Based on the above framework, we divide the whole set of candidate items into multiple semantic clusters. Then we introduce the semantic-aware negative sampling which includes simple negative sampling and hard negative sampling. In simple negative sampling, we select primary clusters for each positive sample based on the following probability formula, ensuring that none of these selected clusters are the same as the primary cluster of the positive sample.
\begin{equation}
    \begin{split}
Q(C_p=i) &= \frac{1}{d(z_p^i, z_p^{+})^\gamma}, \rm{s.t.} ~ i \neq +, \\
P(C_p=i) &= \frac{Q(C_p=i)}{\sum \limits_{j \neq +} Q(C_p=j)},
    \end{split}
    \label{eq:prob}
\end{equation}
where \(d( \cdot, \cdot )\) measures the distance between primary codewords using an inner-product operation, which is subsequently normalized to a range from 0 to 1. \(z_p^{+}\) is the primary cluster of the positive sample, \(Q(C_p=i)\) is the unnormalized sampling probability of similar primary clusters with \(\gamma\), \(P(c_p=i)\) is the normalized sampling probability of primary cluster \(z_p^i\). Then, we randomly select samples from each cluster which enhances the diversity of negative samples. After being encoded by the item tower\ \citep{huang2020embedding}, the embedding set of simple negative samples can be represented as \(V_s\):
\begin{equation}
    \begin{split}
V_s&=\{V_s^1,...,V_s^k,...,V_s^{m_c} \}, \\
V_s^k&=\{\mathbf{v}_s^{(k-1)m_o+1},...,\mathbf{v}_s^{km_o} \},
    \end{split}
    \label{eq:inter}
\end{equation}
where \(V_s^k\) is the embedding set of the simple negative samples in $k$-th cluster, \(m_c\) is the number of selected clusters and \(m_o\) is the number of selected samples in each cluster. In this way, we dynamically adjust the difficulty of the simple negatives as well as  mitigate group-level sampling biases.

In hard negative sampling strategy, we randomly select partially similar samples within the positive primary cluster. Then, we consider samples in the same secondary cluster as false negatives and remove these samples from the hard negative samples set. The output embedding set of hard negative samples can be represented as \(V_h=\{\mathbf{v}_h^1,\mathbf{v}_h^2,...,\mathbf{v}_h^{m_h} \}\), where \(m_h\) is the number of selected samples in the positive primary cluster.

\subsection{Effective Dense Interpolation Strategy}
\label{sec:dins}

By employing our negative sampling process, we obtain semantic clusters and randomly selected negatives from each cluster. It's a well-established principle\ \citep{chen2020simple} that increasing the negative sampling size can enhance the performance of the EBR models. However, the process mentioned above does not guarantee a sufficient sampling size for each cluster. To solve this problem, we propose a parameter-adaptive negative sampling augmentation technique based on the linear interpolation to increase the number of negative samples. The detailed interpolation process is applied to both simple negatives and hard negatives, which is illustrated in Figure \ref{fig:frameworkb}.

\subsubsection{Interpolation on Simple Negative Samples}
Suppose we select \(n_o\) negative anchors (\(2 \leq n_o \leq m_o\)) from the \(k\)-th cluster. The output item embeddings are reordered as \(V_{s_k}=\{\mathbf{v}_{s_k}^{1},...,\mathbf{v}_{s_k}^{n_o} \}\). Each vector in the embedding set is selected once as the anchor vector \(\mathbf{v}_{s_k}^{a}\), and generate the virtual negative samples similar to the embedding set by linear interpolation:
\begin{equation}
\begin{split}
    \tilde{\mathbf{v}}_{s_k}^{a}&=\sum_{i=1}^{n_o}\alpha_i \mathbf{v}_{s_k}^{i}, \\ 
\alpha_i&= \frac{d({\mathbf{v}_{s_k}^{a}},\mathbf{v}_{s_k}^{i})^\eta}{\sum_{j=1}^{n_o}  {d({\mathbf{v}_{s_k}^{a}},\mathbf{v}_{s_k}^{j})^\eta}},
\end{split}
\end{equation}
where \(\tilde{\mathbf{v}}_{s_k}^{a}\) denotes the virtual negative sample obtained by linear interpolation, \(d( \cdot, \cdot )\) is the inner-product operation to measure the embedding distance between item vectors, which is subsequently normalized to a range from 0 to 1. \(\eta\) is designed to adjust the magnitude of impact resulting from surrounding vectors, \(\alpha_j\) is the adaptive parameter to fuse negative samples. Our method ensures that each virtual sample is proximate to the anchor and can be disturbed by other negative samples in terms of similarity. When the number of selected negatives \(n_o\) ranges from 2 to \(m_o\), the quantity of virtual samples \(m_c^v\) in the \(k\)-th primary cluster is proportional to \(O(m_o^2)\), which can be solved as follow:
\begin{equation}
    m_c^v=2+3+...+m_o \varpropto O(m_o^2).
\end{equation}

In this way, we efficiently enhance the diversity and richness of negative samples.

\subsubsection{Interpolation on Hard Negative Samples}

As mentioned in the previous explanation, the hard negative samples are selected from the same primary cluster but different secondary cluster. The interpolation on hard negative samples is proposed to further augment the quantity of samples and facilitate the controllable difficulty of hard negative samples. We denote the output embedding of positive sample as \(\mathbf{v}_{+}\). We conduct the linear interpolation between \(\mathbf{v}_{+}\) and each existing hard negative sample \(\mathbf{v}_{h}^a\):
\begin{equation}
    \tilde{\mathbf{v}}_{h}^{a}=\lambda  \mathbf{v}_{+}+(1-\lambda)\mathbf{v}_{h}^a,
\label{eq:hard_inter}
\end{equation}
where \(\lambda\) is a hyperparameter used to adjust the difficulty of hard negatives during the training process. When \(0 < \lambda < 1\), virtual hard negatives are generated between positive samples and existing hard nagetives which provides more challenging samples. When \(\lambda < 0\), virtual hard negatives are easier than existing hard nagetives which provides relatively simple samples. By this strategy, we enhance the challenge of discriminating the classification boundary and incorporate the stochastic uncertainty into the model which also improves its generalization performance.

\subsection{Model Learning}
\label{sec:ml}
Following the widely used EBR method, Deep Structured Semantic Model (DSSM)\ \citep{wang2018additive, dssm}, we can optimize the similarity between user embeddings \(\mathbf{u}\) and item embeddings \(\mathbf{v}\) by contrastive learning method. The objective function applied is the InfoNCE loss, which is defined as follows:
\begin{equation}
\mathcal{ L}_{sm}=- \frac {1 }{N} \sum_{i=1}^{N}\log \frac {e^{\mathbf{u}_i^T \mathbf{v}_{i}^+ / \tau}} {e^{\mathbf{u}_i^T \mathbf{v}_{i}^{+} / \tau} + \sum_{i=1}^{N_s} e^{\mathbf{u}_{j}^T \mathbf{v}_{i}^{j-} / \tau}},
\label{eq:ams}
\end{equation}
where \(N_s\) is the number of negatives effectively sampled by our ESANS method, \(\mathbf{v}_i^{j-}\) denotes the \(j\)-th negative sample of the \(i\)-th positive sample \(\mathbf{v}_i^{+}\), \(\mathbf{u}^T\mathbf{v}\) is also called as the target logit\ \citep{pereyra2017regularizing} of the \(i\)-th positive sample, \(\tau\) is the temperature hyperparameter used to adjust the distribution of logits. \textbf{The algorithm process is presented in Appendix \ref{appd:Pseudocode}}.




\section{Offline Experiments}
In this section, we conduct offline experiments on three real-world datasets to demonstrate the effectiveness and efficiency of our proposed method. The first two are public datasets, while the third is an in-house industrial dataset. The descriptions and statistics of the two public datasets and the industrial dataset are detailed in Table \ref{dataset2}, respectively. Additionally, we perform an ablation study of our modules and address the following research questions:
\begin{itemize}[noitemsep, topsep=0pt, leftmargin=*]
\item \textbf{RQ1}: How does our ESANS perform compared to other state-of-the-art models?
\item \textbf{RQ2}: What is the impact of each component on the overall model's performance?
\item \textbf{RQ3}: What is the effect of the hyper-parameters on the performance of our model?
\end{itemize}
\vspace{-3pt}
\begin{table}[htbp]
    \centering
    \caption{Statistics of Public and Industrial Datasets.}
    \label{dataset2}
    \vspace{-10pt}  
    \resizebox{230pt}{!}{  
        \begin{tabular}{c|c|c|c|c|c|c}
            \toprule
            Dataset & Amazon Elecs & Pixel-Rec & \#A1 & \#A2 & \#A3 & \#A4 \\
            \midrule
            \#User & 247,446 & 29,845,039 & 4,930,611 & 24,931,581 & 17,037,221 & 15,914,765 \\
            \#Item & 88,408 & 408,374 & 2,163,338 & 4,268,324 & 2,905,716 & 3,067,253 \\
            \#Interaction & 2,146,317 & 195,755,320 & 61,579,472 & 336,744,161 & 149,341,806 & 163,611,291 \\
            \bottomrule
        \end{tabular}
    }
    \vspace{-18pt}
\end{table}

\subsection{Experimental Setup}
\textbf{Dataset.}
\begin{itemize}[noitemsep, topsep=0pt, leftmargin=*]
\item \textbf{Amazon Review.} It was first introduced by Van Gysel et al.\ \citep{van2016learning,van2017semantic} and has become a benchmark dataset for evaluating product recommendation methods\ \citep{fan2022modeling, tang2022croloss, ma2023exploring}. We select the Electronics subset which products a sufficient number of user reviews and includes comprehensive metadata, such as product titles and categories. The textual features are extracted by sentence-transformers\ \citep{reimers2019sentence} from \ \citep{zhou2023bootstrap} and the visual features are extracted and published in \ \citep{ni2019justifying}.

\item \textbf{Pixel-Rec.} This dataset\ \citep{cheng2024image} is derived from a global online video platform which captures approximately 200 million user consumption from September 2021 to October 2022. It focuses on content-driven recommendations spanning diverse categories such as food, games, fashion, and makeup. The textual and visual features of these contents have already been extracted using PixelNet, a network proposed concurrently with Pixel-Rec.

\item \textbf{Industrial Dataset.} We establish the in-house offline dataset by collecting the users' sequential behaviors and feedback logs from Alibaba’s international e-commerce platform, Lazada. The dataset comprises four categories, each representing a distinct Southeast Asian country, labeled from \#A1 to \#A4. 
\end{itemize}

\noindent\textbf{Graph Construction.}
Due to space limitation, the introduction of behavior-based graph construction is provided in Appendix \ref{appd:graph}. 

\begin{table*}[htbp!]
\centering
\caption{Performance Comparison across baselines. The last column (AVG) denotes the average improvement of sampling methods across all datasets. The last row (RI) denotes the relative improvement of our ESANS over UNS.}
\resizebox{\linewidth}{!}{
    \begin{tabular}{c|cc|cc|cc|cc|cc|cc|cc}
    \toprule
    \multirow{2}{*}{\textbf{Method}} & \multicolumn{2}{c|}{\textbf{Amazon Elecs}} & \multicolumn{2}{c|}{\textbf{Pixel-Rec}} & \multicolumn{2}{c|}{\textbf{\#A1}} & \multicolumn{2}{c|}{\textbf{\#A2}} & \multicolumn{2}{c|}
    {\textbf{\#A3}} & \multicolumn{2}{c|}{\textbf{\#A4}} & \multicolumn{2}{c}
    {\textbf{AVG}}\\
     & \centering Recall@50 & \centering Recall@200 & Recall@50 & Recall@200 & Recall@50 & Recall@200 & Recall@50 & Recall@200 & Recall@50 & Recall@200 & Recall@50 & Recall@200 & Recall@50 & Recall@200 \\ 
    \midrule
    UNS &0.1633&0.4512&0.0737&0.1522&0.4276& 0.6087&0.3108&0.5155&0.3859&0.6478&0.3857&0.6277&0.2912&0.5005\\ 
    PNS &0.1695&0.4696&0.0783&0.1587&0.4332&0.6401&0.3370&0.5318&0.3803&0.6405&0.3883&0.6321&0.2978&0.5121\\ 
    debiased MNS &0.1874&0.4726&0.0801&0.1655&0.4549&0.6562&0.3597&0.5647&0.3970&0.6513&\underline{0.4074}&\underline{0.6549}&0.3144&0.5275\\ 
    MixGCF &0.1963&0.4759&0.0794&0.1631&0.4593&0.6577&0.3621&0.5703&0.4012&0.6574&0.3924&0.6483&0.3151&0.5288\\ 
    FairNeg &0.1792&0.4705&\underline{0.0836}&\underline{0.1782}&\underline{0.4790}&0.6688&\underline{0.3669}&\underline{0.6052}&0.4043&\underline{0.6687}&0.3983&0.6501&0.3186&\underline{0.5403}\\ 
    Adap-$\tau$ &\underline{0.2018}&\underline{0.4873}&0.0763&0.1684&0.4694&\underline{0.6757}&0.3538&0.5883&\underline{0.4097}&0.6625&0.4046&0.6437&\underline{0.3193}&0.5377\\ 
    \textbf{ESANS (ours)} &\textbf{0.2135}&\textbf{0.4948}&\textbf{0.0908}&\textbf{0.1828}&\textbf{0.4862}&\textbf{0.6918}&\textbf{0.3887}&\textbf{0.6216}&\textbf{0.4176}&\textbf{0.6732}&\textbf{0.4182}&\textbf{0.6609}&\textbf{0.3358}&\textbf{0.5542}\\ \midrule
    RI &+30.74\%&+9.66\%&+23.20\%&+20.11\%&+13.70\%&+13.65\%&+25.06\%&+20.58\%&+8.21\%&+3.92\%&+8.43\%&+5.29\%&+15.32\%&+10.73\%\\
    \bottomrule
    \end{tabular}
}
\label{table:results}
\end{table*}

\noindent \textbf{Baselines.}
We compared our ESANS with five representative negative sampling methods based on the classical two-tower architecture. The methods are as follows:
\begin{itemize}[noitemsep, topsep=0pt, leftmargin=*]
\item \textbf{UNS}\ \citep{rendle2012bpr, he2017neural}: A widely used negative sampling approach involves randomly selecting instances from a uniform distribution.
\item \textbf{PNS}\ \citep{chen2017sampling}: A negative sampling method that adjusts the sampling distribution based on item popularity.
\item \textbf{Debiased MNS}\ \citep{yi2019sampling,yang2020mixed}: A method that integrates UNS with in-batch negative sampling, and introduces a technique to address the oversampling issue of popular items.
\item \textbf{MixGCF}\ \citep{huang2021mixgcf}: A method synthesizes hard negatives between negatives and positives in a graph-based model. We adapt this to a two-tower structure to generate virtual hard negatives in the item representation space.

\item \textbf{FairNeg}\ \citep{chen2023fairly}: A method that improves item group fairness by adaptively adjusting the distribution of negative samples at the group level.
\item \textbf{Adap-\(\tau\)}\ \citep{chen2023adap}: A method that adjusts the temperature coefficient of the loss function by the embedding similarity between users and corresponding negatives.
\end{itemize}

\noindent \textbf{Evaluation Metrics.}
For the evaluation metrics in recommendation tasks, we follow\ \citep{bansal2016ask, wang2015collaborative} and use Recall@K for each group based on the Top-K recommendation results. Finally, the Recall@K is averaged over all users.

\noindent \textbf{Parameter settings.}
Due to space limitation, the implementation details are provided in Appendix \ref{appd:ps}.

Table \ref{table:results} summarizes the overall performance of our ESANS as well as the baselines on both industrial and public datasets, with the best results emphasized in bold and the second-best results underlined. It is noteworthy that ESANS consistently outperforms all baseline methods across the aforementioned datasets, achieving an average improvement of up to \textbf{15.32\%} in Recall@50 and \textbf{10.73\%} in Recall@200 compared to its base method UNS. PNS generally outperforms UNS across most datasets, indicating that boosting the sampling possibility for popular items improves sampling quality. However, it is worth noting that PNS does not exceed UNS performance in the \#A3 dataset, which might be attributed to the introduced popularity bias. Once the challenge of popularity bias is addressed, the debiased MNS Sampling method outperforms UNS and PNS across all datasets and outperforms other baselines on \#A4. MixGCF introduces virtual hard negatives by hop-mixing interpolation which achieves similar performance with the debiased MNS and proves the feasibility of hard negatives augmentation. However, the interpolation process fails to consider semantics and yields noisy negatives, so it is outperformed by our method. FairNeg is another work conducted to reduce the sampling bias via adjusting the group-level negative sampling distribution which provides the best recommendation utility on Pixel-Rec and \#A2 in all baselines. However, this work determines the groups by the only item attribute view which is not comprehensive and thus is surpassed by our method. Ada-\(\tau\) is proposed to design a learnable \(\tau\), which enables the adaptive adjustment of the difficulty level for negatives. This work outperforms other baseline models on Amazon Elecs. However, Ada-\(\tau\) fails to provide incremental information by deriving more challenging negatives so that it is beaten by our method. 

In summary, our method effectively addresses the inherent limitations of these methods and achieves SOTA performance across all datasets in terms of retrieval efficiency. It is worth noting that the MSAC module is actually detached from the EBR model's training process and the EDIS module is only applied to the output embeddings of the EBR model. Therefore, our method does not introduce additional computational complexity for offline training. The comparison of time costs among these methods in the training process is shown in Table \ref{table:time_cost}, which strongly supports our statement. Furthermore, ESANS can be deployed online similarly to other classical EBR models, with the user and item towers deployed separately. This ensures that the online service costs remain comparable to those of other baseline models.

\begin{table}[htbp]
    \centering
    \caption{Time cost in EBR training process on the \#A2 Dataset.}
    \label{table:time_cost}
    \vspace{-5pt}
    \resizebox{240pt}{!}{
    \begin{tabular}{c|c|c|c|c|c|c|c}
        \toprule
        \multirow{1}{*}{Method} & \multicolumn{1}{c}{UNS} & \multicolumn{1}{c}{PNS} & \multicolumn{1}{c}{Debiased MNS} & \multicolumn{1}{c}{MixGCF} & \multicolumn{1}{c}{FairNeg} & \multicolumn{1}{c}{Adap-$\tau$} & \multicolumn{1}{c}{ESANS} \\
        \midrule
        Time Costs           & \multicolumn{1}{c}{9h15min} & \multicolumn{1}{c}{9h23min} & \multicolumn{1}{c}{10h32min} & \multicolumn{1}{c}{9h47min} & \multicolumn{1}{c}{14h46min} & \multicolumn{1}{c}{13h9min} & \multicolumn{1}{c}{10h56min} \\
        \bottomrule
    \end{tabular}
    }
\vspace{-18pt}
\end{table}

\subsection{Ablation Study (RQ2)}

To investigate the effectiveness of each component in the proposed model, in this subsection, we conduct a series of ablation studies on the \#A2 industrial dataset, as it represents the most complex and representative scenario with the largest user scale and the richest behavior on our platform. The specific experiment settings are introduced as follows:
\begin{itemize}[noitemsep, topsep=0pt, leftmargin=*]
\item \textbf{w/o MSAC}, removes the Multimodal Semantic-Aware Clustering before the Interpolation-based negative sampling.
\item \textbf{w/o EDIS}, removes the Effective Dense Interpolation Strategy employed in both simple negative sampling and hard negative sampling strategies. Furthermore, we conduct additional ablation studies on both simple and complex interpolation strategies.
\item \textbf{w/o Multimodal Aligning}, removes the textual and visual modalities and reserves the behavior-based modality for further clustering. Considering the relatively high cost of using three modals, we also remove each of the three modals to evaluate their individual contributions.
\item \textbf{w/o Secondary Codebook}, removes the secondary codebook in the Vector Quantized Clustering, thereby invalidating the interpolation-based hard negative sampling.
\end{itemize}
\begin{table}[htbp!]
\caption{Ablation Study on the \#A2 Dataset.}
\centering
\resizebox{220pt}{!}{
\begin{tabular}{l|cc}
\toprule
\multirow{2}{*}{\centering Method} & \multicolumn{2}{c}{\#A2} \\ 
 & Recall@50 & Recall@200 \\ 
 \midrule
\textbf{Ours}
&\textbf{0.3887}&\textbf{0.6216}\\ 
w/o MSAC &0.3653&0.5976\\ 
w/o EDIS &0.3788&0.6134\\
\footnotesize {\hspace{5pt} ---w/o simple interpolation} &\footnotesize {0.3865}&\footnotesize {0.6187}\\ 
\footnotesize {\hspace{5pt} ---w/o hard interpolation} &\footnotesize {0.3822}&\footnotesize {0.6169}\\ 
w/o Multimodal Aligning &0.3802&0.6163\\ 
\footnotesize {\hspace{5pt} ---w/o visual modal} &\footnotesize {0.3857}&\footnotesize {0.6191}\\ 
\footnotesize {\hspace{5pt} ---w/o textual modal} &\footnotesize {0.3848}&\footnotesize {0.6174}\\ 
\footnotesize {\hspace{5pt} ---w/o behavior-based modal} &\footnotesize {0.3786}&\footnotesize {0.6159}\\ 
w/o Secondary Codebook &0.3724&0.6082\\ 
\bottomrule
\end{tabular}
}
\label{table:ablation}
\end{table}
\setlength{\textfloatsep}{15pt} 

Table \ref{table:ablation} presents the performance of these ablation experiments. Firstly, we can observe that adopting Multimodal-aligned Clustering Algorithm improves Recall@50 by \textbf{6.41\%} and Recall@200 by \textbf{4.02\%}, which proves that the semantic clustering algorithm employed for interpolation significantly improves sampling quality. The dense interpolation respectively brings a \textbf{2.61\%} and a \textbf{1.34\%} improvement for Recall@50 and Recall@200, which demonstrates the efficient of our sample augment strategy. We also find that hard interpolation achieves more improvement compared with simple interpolation, which implies the importance of hard negatives. Besides, the Multimodal clustering performs better than the Unimodal clustering (\textbf{2.24\%} on Recall@50 and \textbf{0.86\%} on Recall@200), which superiority the multi-view representations. Among the three modals, the behavior-based modality exhibits the best performance, while the visual and textual modalities also contribute positively to the MSAC. The interpolation-based hard negative sampling conducted by the secondary codebook also shows the improvement (\textbf{4.38\%} on Recall@50 and \textbf{2.20\%} on Recall@200), which proves the feasibility of selecting hard negative samplings with heuristics semantic constraint.
\vspace{-0.6cm}
\subsection{Hyperparameters Sensitivity Analysis (RQ3)}
\begin{figure}
  \includegraphics[width=0.46\textwidth]{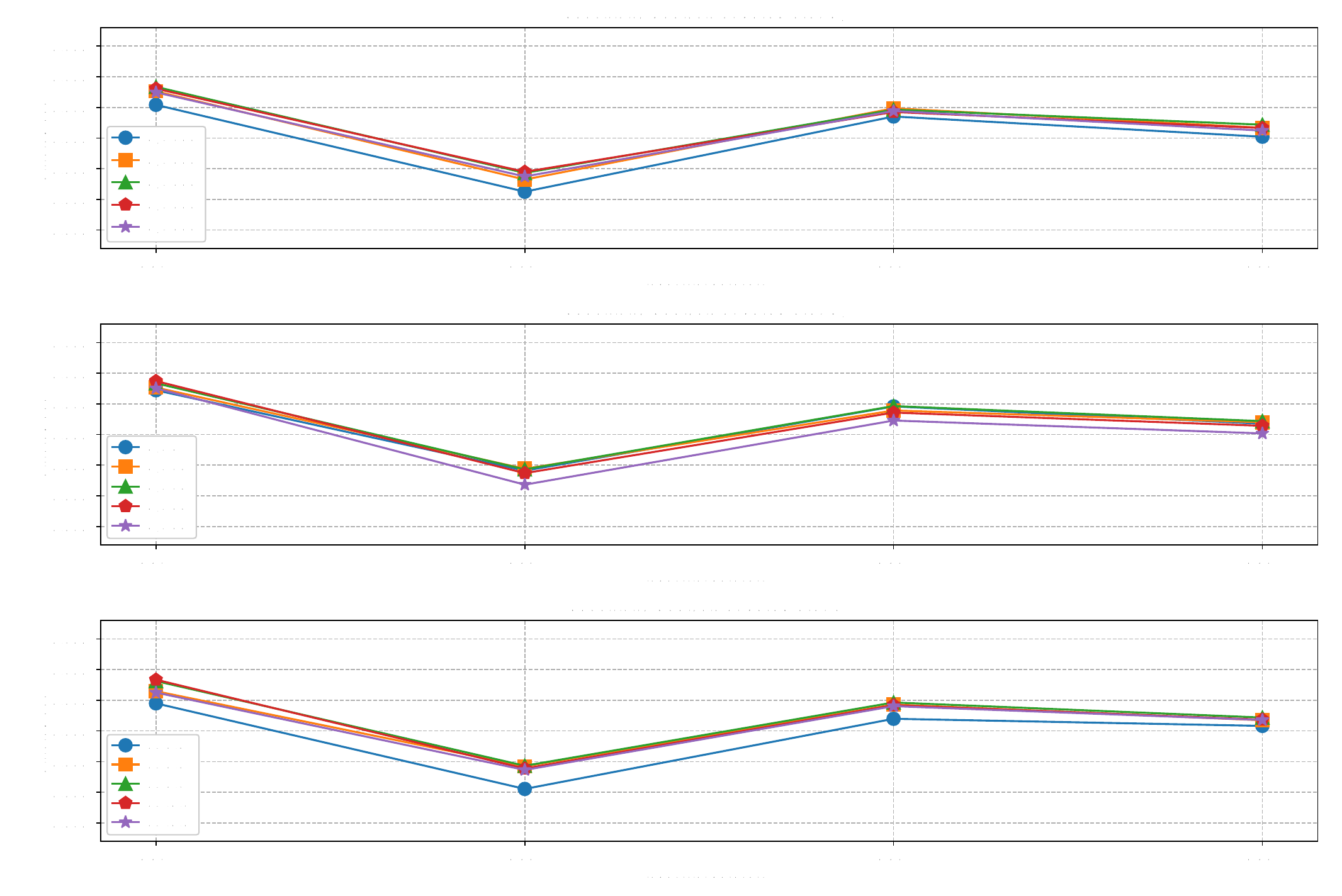}
  \caption{The performance of ESANS under different hyper-parameters(\(K_p\), \(K_s\) and \(\lambda\)) on \#A1-\#A4 industrial datasets.}
  \Description{xx.}
  \label{fig:sensitive_v1}
  \vspace{-3pt}
\end{figure}

In this section, we investigate the sensitivity of our model's hyperparameters, specifically the number of primary clusters \(K_p\), the number of secondary clusters \(K_s\) and the interpolation coefficient \(\lambda\). These experiments are carried out on the \#A1-\#A4 industrial datasets, employing five distinct values for \(K_p\) (100, 200, 300, 400, 500), \(K_s\) (5, 10, 15, 20, 25) and \(\lambda\) (-0.3, -0.1, 0.1, 0.3, 0.5).  Figure \ref{fig:sensitive_v1} illustrates the performance of these hyperparameter tuning experiments. We observe that the model's performance stays consistently high when \(K_p\) is increased from 200 to 500. However, reducing \(K_p\) to 100 leads to a slight decrease in performance. This observation encourages us to consider a higher value for \(K_p\) to further enhance the intra-cluster semantic consistency. \(K_s\) shows optimal prediction performance between 5 to 15. We recommend to set a relatively small value for \(K_s\) in order to minimize the occurrence of false negatives. According to our experiments on \(\lambda\), \(\lambda=0.1\) achives the best performance on \#A2-\#A4 while \(\lambda=-0.1\) achives the best performance on \#A1. We find that \(\lambda\) is relatively sensitive. As we increase \(\lambda\) to 0.5, the performance across all datasets declined significantly. This is due to the fact that the generated virtual false negatives are very close to the positives, which may confuse the model. It is also worth noting that sometimes \(\lambda\) can be set to less than 0 to achieve better performance, which suggests that introducing easier hard negatives may also be helpful.

\section{Online Experiments}
To further validate the effectiveness of our approach, we conducted an online A/B test on an e-commerce recommendation platform from September 13 to 19, 2024. The control group used a two-tower model with debiased mixed negative sampling (MNS)\ \citep{yang2020mixed}, while the experiment group applied our proposed method. Both groups consisted of 30\% randomly selected users. Specifically, we observed \textbf{2.83\%} increase in the \textbf{Advertising Revenue}, \textbf{1.19\%} increase in the \textbf{Click-Through-Rate(CTR)} and \textbf{1.94\%} increase in the \textbf{Gross Merchandise Volume(GMV)}. The results of the online experiment once again confirm the efficiency and effectiveness of our method ESANS in negative sampling for recommendation systems.

\section{Conclusion}
In this study, we proposed a novel negative sampling method, Effective and Semantic-Aware Negative Sampling (ESANS), which integrates an Effective Dense Interpolation Strategy (EDIS) and Multimodal Semantic-Aware Clustering (MSAC). Extensive experiments demonstrated that ESANS significantly improves sampling quality and efficiency compared to baselines. Specifically, EDIS improves the diversity and density of the sampling distribution. MSAC enhances semantic consistency and reduces false negatives. These modules advance the effectiveness of negative sampling in recommendation systems. For future work, we will pursue two directions. The first is to further optimize the multimodal representations based on MSAC. The second direction is to design a more complex interpolation strategy among the outputs of hidden layers.






\balance

\bibliographystyle{acm-reference-format}
\bibliography{main}

\appendix

\section{Algorithm}
\vspace{-3pt}
\label{appd:Pseudocode}
\begin{figure}[htbp!]
  \includegraphics[width=0.43\textwidth]{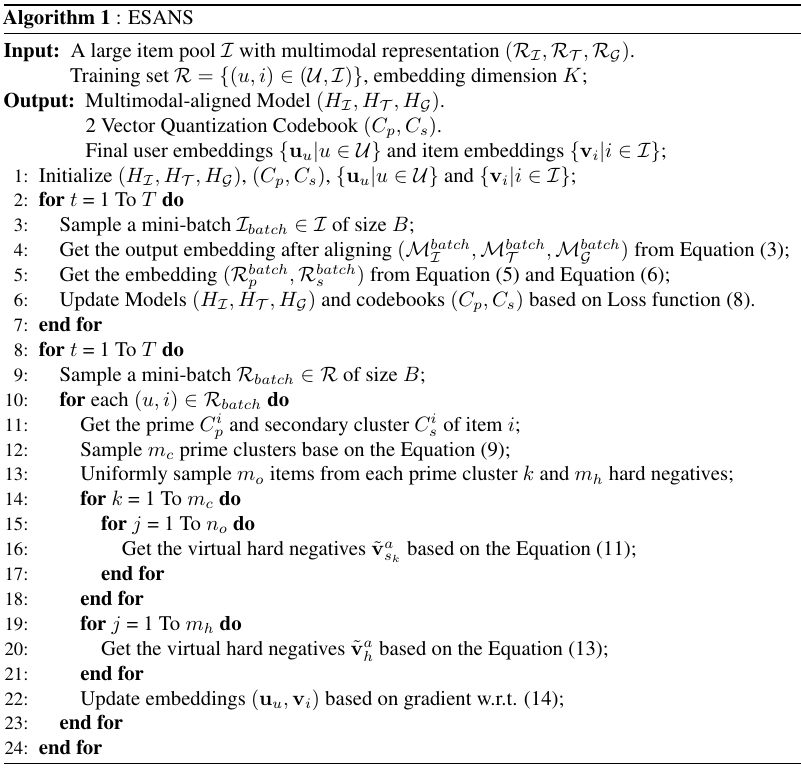}
  \Description{xx.}
  \label{fig:code}
\vspace{-8pt}
\end{figure}

\section{Graph Construction}
\label{appd:graph}
For each dataset, we pretrain a heterogeneous graph network\ \citep{li2022soft} based on user behaviors. The types of graph nodes include user, item, and its side information (brand / category / price features for Amazon Review dataset, tag / statistical features for Pixel-Rec dataset and brand / shop / category for industrial datasets). The graph edges include: 1) user-item edge. If user \(u\) clicks item \(i\), there is an edge between \(u\) and \(i\). 2) user-side information edge. If user \(u\) clicks an item with side information \(v\) (e.g., shop), there is an edge between \(u\) and \(v\). 3) item-item edge. If item \(i\) and item \(j\) are adjacent in user behavior sequence and the time interval between item \(i\) and item \(j\) is within 60 seconds, there is an edge between \(i\) and \(j\). 4) item-side information edge. If item \(i\) has a side info \(v\), there is an edge between \(i\) and \(v\).

\section{Parameter Settings}
\label{appd:ps}
In this section, we elaborate on the parameter settings for the implementation of our algorithm.
To ensure computational manageability, we limit the length of user behavior sequences to 10 for the Amazon Review dataset, 32 for the Pixel-Rec dataset and 64 for the Industrial dataset. The training process is implemented using a distributed TensorFlow\cite{tensorflow2016abadi} platform, consisting of 10 parameter servers and 40 workers with 12 CPUs per worker. It is worth noting that the performance of our method can be further enhanced as the sampling scale increases, as shown in Table \ref{table:mc_mo}. To ensure fairness between our ESANS and the baseline models, in the negative sampling process, for each in-batch positive sample, we randomly select {\(m_c=2\)} clusters and then draw {\(m_o=5\)} negative samples from each of these clusters. In contrast, the baseline models select 10 negative samples randomly for each positive sample. These negatives are sampled based on an online sampling framework in the training process and shared across the batch. Additionally, the interpolation coefficient \(\lambda\) of hard negatives is set to 0.1 for harder interpolation and -0.1 for easier interpolation. The rest of the hyperparameters settings are demonstrated in Table \ref{Hyper_v2}.

\begin{table}[htbp!]
\caption{Sampling scale experiments for ESANS on \#A2.}
\vspace{-8pt}
\centering
\resizebox{150pt}{!}{
\begin{tabular}{l|cc}
\toprule
\multirow{2}{*}{\centering Sampling Scale} & \multicolumn{2}{c}{\#A2} \\ 
 & Recall@50 & Recall@200 \\ 
 \midrule
\textbf{$m_c=2$, $m_o=5$}
&\textbf{0.3887}&\textbf{0.6216}\\ 
\textbf{$m_c=2$, $m_o=4$} &0.3875&0.6197\\ 
\textbf{$m_c=2$, $m_o=6$} &0.3899&0.6232\\
\textbf{$m_c=1$, $m_o=5$} &0.3793&0.6164\\ 
\textbf{$m_c=3$, $m_o=5$} &0.3930&0.6241\\ 
\bottomrule
\end{tabular}
}
\label{table:mc_mo}
\vspace{-15pt}
\end{table}

\begin{table}[b!]
    \centering
    \caption{Hyper-parameter settings of Our ESANS.}
    \vspace{-8pt}  
{  
    \resizebox{200pt}{!}{
        \begin{tabular}{c|c}
            \toprule
            Hyper-parameter & Choice\\
            \midrule
            \(B\)& 512\\
            \(\tau\) & 0.05 \\
            \(d_k\)& 64 \\
            \(d_m\)& 512 \\
            \(K_p\)& 300 \\
            \(K_s\)& 15 \\
            \(\eta\)& 0.6 \\
            \(\beta_1\ , \beta_2\ , \beta_3\)& 2.0 \\
            Optimizer & Adam\\
            Learning rate & 0.0002\\
            Amazon modal emb size& Img: 4096, Text:384, Graph:128 \\
            PixelRec modal emb size& Img: 1024, Text:1024, Graph:1024 \\
            \#A1-\#A4 modal emb size& Img: 1024, Text:1024, Graph:1024 \\
            \bottomrule
        \end{tabular}
    }
}
\label{Hyper_v2}
\end{table}

\end{document}